\documentclass[fleqn,usenatbib,onecolumn]{mnras}

\usepackage[T1]{fontenc}
\usepackage{graphicx}	
\usepackage{amsmath}	
\usepackage{amssymb}	

\usepackage{newtxtext,newtxmath}

\newcommand{\p}{\partial}
\newcommand\ra{r_{A}}
\newcommand\rb{r_{B}}
\newcommand\rc{r_{C}}
\newcommand\bra{\mathbf{r}_{A}}
\newcommand\brb{\mathbf{r}_{B}}
\newcommand\brc{\mathbf{r}_{C}}

\newcommand\half{\tfrac{1}{2}}
\newcommand\msun{M_\odot}
\newcommand\dy{\mbox{\,d}}
\newcommand\yr{\mbox{\,yr}}
\newcommand\pc{\mbox{\,pc}}
\newcommand\Gyr{\mbox{\,Gyr}}
\newcommand\Myr{\mbox{\,Myr}}
\newcommand\comment[1]{}

\title[Resonant capture in quadruple systems]{Resonant capture in quadruple stellar systems}

\author[S.\ Tremaine]{Scott Tremaine$^{1}$\thanks{tremaine@ias.edu}
\\
$^{1}$Institute for Advanced Study, Princeton, NJ 08540, USA
}

\pubyear{2020}

\begin{document}
\label{firstpage}
\pagerange{\pageref{firstpage}--\pageref{lastpage}}
\maketitle

\begin{abstract}
  Some quadruple star systems in the hierarchical 2+2 configuration exhibit orbit-orbit resonances between the two compact binaries. We show that the most important resonances occur at period ratios of 1:1, 3:2 and 2:1. We describe the conditions required for capture and show that they can be satisfied at the 3:2 and 2:1 resonances in binaries that migrate significantly in semimajor axis after circularization, probably through magnetic braking or gravitational radiation. 
\end{abstract}


\section{Introduction}

\medskip

Quadruple stellar systems are almost always found in hierarchical configurations, since this is the only simple way to ensure long-term orbital stability. There are two possible architectures: the 3+1 configuration, in which a hierarchical triple system is orbited by an even more distant companion, and the 2+2 configuration, in which two relatively compact binaries orbit their common center of mass. In this paper we explore novel resonant phenomena that can arise in the 2+2 configuration.

The best-studied 2+2 systems are those in which both compact binaries are eclipsing. These `doubly eclipsing' systems are rare: the most recent compilation \citep{zasche19} lists only 146 systems, and a significant fraction of these may be superimposed pairs of unrelated eclipsing binaries.

\begin{figure} \includegraphics[width=0.95\textwidth]{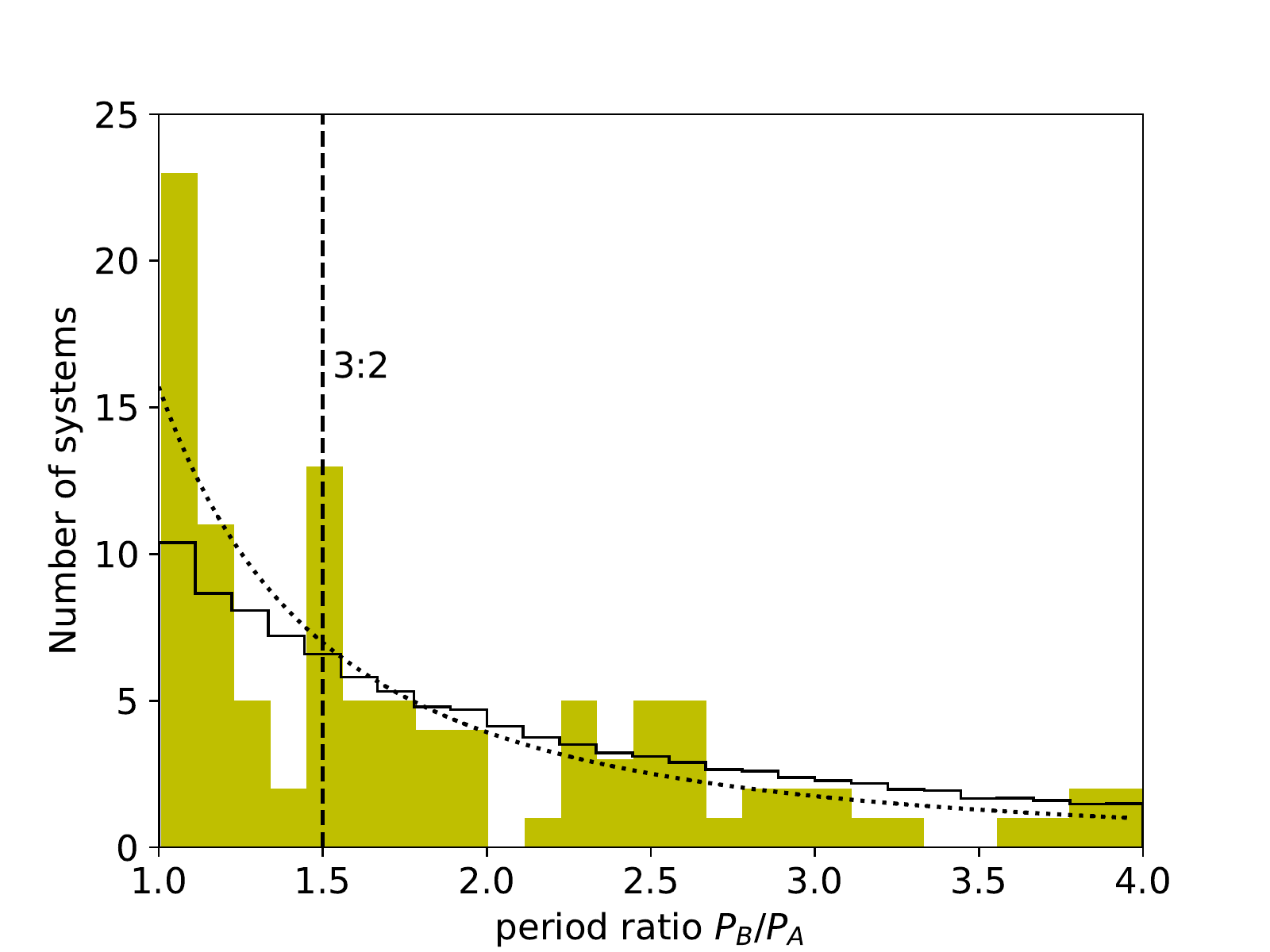} \caption{Period ratios of doubly eclipsing 2+2 quadruple star systems, from the compilation of \citet[][compare their Figure 13]{zasche19}.  There are peaks at the period ratios 1:1 and 3:2. The black histogram shows the period-ratio distribution obtained by randomly selecting pairs of binaries from the quadruple systems in this sample. The black dotted curve shows the period-ratio distribution $\propto (P_B/P_A)^{-2}$ expected if the binary periods are drawn at random from a uniform distribution. Neither the histogram nor the curve reproduces the strength of the peak at period ratio  1:1.}
\label{fig:one}
\end{figure}

Despite the modest sample size, \cite{zasche19} find significant peaks in the distribution of period ratios $P_B/P_A$ near 1 and 1.5 (see Figure \ref{fig:one}). There is no evidence for peaks near other resonances such as $P_B/P_A=2.$

Here are more complete descriptions of a few of the near-resonant systems from this sample. 

\begin{itemize}

\item V994 Her \citep{lee08,zasche16} consists of two detached eclipsing binary systems with periods $P_A=1.4200\dy$, $P_B=2.0833\dy$. (Here and throughout the paper, the labels `A' and `B' refer to the binaries with the shorter and longer periods, and `C' refers to the mutual orbit of the centers of mass of binaries A and B. Not all authors follow this convention.)  The distance is $320\pm10\pc$ based on the Gaia DR2 parallax of $3.05\pm0.13\mbox{\,mas}$ and an assumed parallax offset of $0.05\mbox{\,mas}$. System A has spectral types A2V and A4V, while B has spectral types B8V and A0V. The stellar masses range from $1.8M_\odot$ to $3.0M_\odot$; the main-sequence lifetime of the most massive star is about $3\times10^8\yr$, suggesting that the age of the system is less than this. Both systems have non-zero eccentricities: $e_A=0.125\pm0.007$ and $e_B=0.031\pm0.001$. The mutual orbit has period $P_C=2.91\pm0.14\yr$ and eccentricity $e_C=0.76\pm0.07$. The period ratio is close to but not exactly 3:2, with $1-2P_B/3P_A=0.0220$.

\item OGLE BLG-ECL-145467 \citep{zasche19} has $P_A=3.305\dy$, $P_B=4.910\dy$, $P_C= 1540\pm90 \dy$. The distance is  $3100\pm400\pc$ based on the Gaia DR2 parallax of $0.374\pm0.048\mbox{\,mas}$ and the same assumed parallax offset. All three orbits have non-zero eccentricity, $e_A=0.007\pm0.001$, $e_B=0.068\pm0.003$, and $e_C=0.395\pm0.018$. Apsidal motion is detected in binary A with $2\upi/\dot\varpi\simeq 125\yr$. The stars appear to have spectral types near B9. The period ratio is close to a 3:2 resonance,  with $1-2P_B/3P_A=0.00961$.

\item OGLE LMC-ECL-10429 or BI 108 \citep{kol13} has $P_A=3.577936\dy$ and $P_B=5.366616\dy$. The flux from binary B is dominated by two B1 supergiants.  The period ratio is close to a 3:2 resonance, with $1-2P_B/3P_A=0.000054$.

\item CzeV343 \citep{cp12} is a doubly eclipsing system with $P_A=0.8069\dy$ and $P_B=1.2094\dy$. The eccentricity of binary A is indistinguishable from zero and $e_B=0.18$. Isolated eclipsing binaries with this period typically have smaller eccentricities. The period ratio is remarkably close to 3:2, with $1-2P_B/3P_A=(8.5\pm0.2)\times10^{-4}$. No parameters for the mutual or C orbit are available and the only direct evidence that the system is not a pair of unrelated binaries is the near-resonance in the periods. 

\item KIC 4247791 \citep{leh12} consists of two eclipsing binaries with periods $P_A=4.0497\dy$ and $P_B=4.1009\dy$ at a distance of $1180\pm90\pc$. Both binaries have small $(<0.006$) but non-zero orbital eccentricities. System A has spectral types F0IV and F2IV, while B has spectral types F7V and F8V. The period ratio is close to 1:1, with $1-P_B/P_A=-0.0126$.

\end{itemize}

The aim of this paper is to investigate the dynamics of resonant capture in 2+2 quadruple stars. The theory of resonant capture was developed for satellites of the solar system \citep{gold65,yoder79,henrard82,hl83,gs14} and most of our techniques are borrowed from that subject. Resonant capture in quadruple stellar systems was examined by \cite{bv18}; they restricted their attention to capture into the 1:1 resonance and found that capture was unlikely, a conclusion that we confirm. \cite{seto18} considered capture into the 1:1 resonance in a quadruple system emitting gravitational radiation.

Most of the longer calculations in this paper are relegated to appendices. 

\bigskip

\section{Preliminaries}

\medskip

\subsection{Notation}

\smallskip

We examine a quadruple system consisting of two binary subsystems A and B. Binary A contains two stars with masses $m_1$ and $m_2$ and binary B contains stars with masses $m_3$ and $m_4$. The reduced and total mass in each binary are denoted by
\begin{equation}
    \mu_A\equiv \frac{m_1m_2}{m_1+m_2}, \quad M_A\equiv m_1+m_2, \quad  \mu_B\equiv \frac{m_3m_4}{m_3+m_4}, \quad M_B\equiv m_3+m_4.
\end{equation}
The position vector $\bra$ is directed from $m_2$ to $m_1$ and the vector $\brb$ points from $m_4$ to $m_3$. We use the term `binary C' to refer to the fictitious system consisting of point masses $M_A$ and $M_B$ located at the centers of mass of binaries A and B. The separation of binary C is denoted by $\brc$, which points from the center of mass of binary B to the center of mass of binary A.

We are interested in the case where all four stellar masses $m_1,\ldots,m_4$ are similar, the binary semimajor axes $a_A$ and $a_B$ are similar, and the quadruple system is hierarchical, that is, there is a small parameter 
\begin{equation}
    \epsilon =\max(a_A,a_B)/a_C \ll 1.
\label{eq:epsdef}
  \end{equation}
  Assuming that all four stars have the same mass this parameter can be expressed in terms of the orbital periods,
  \begin{equation}
    \epsilon=2^{-1/3} \left[\frac{\max(P_A,P_B)}{P_C}\right]^{2/3}.
    \label{eq:epsdefa}
  \end{equation}
  For the 12 quadruple systems with the best determination of all three orbital periods \citep[][Table 4]{zasche19}, $\epsilon$ ranges from 0.034 to 0.0068 with a median of 0.011. Selection effects disfavor even smaller values of $\epsilon$ because $P_C$ must typically be shorter than the data span.  Selection effects also disfavor larger values of $P_A$ and $P_B$ because short-period binaries are more likely to be eclipsing. 

  We show below that the relevant resonant term in the Hamiltonian is $\mbox{O}(\epsilon^5)$. The ability of these relatively weak resonances to capture a tidally evolving binary arises because there is a second small parameter in the system, the ratio of the orbital periods of the compact binaries $P_A, P_B$ to the tidal evolution time $\tau$ (\S\ref{sec:tidal}). 

The osculating Kepler orbital elements for binary A include the semimajor axis $a_A$, the eccentricity $e_A$, the mean longitude $\varpi_A$, and the mean longitude $\lambda_A$. The mean motion is $n_A=(GM_A/a_A^3)^{1/2}$. We know that the angular-momentum vectors of binaries A and B must lie approximately in the sky plane since both binaries eclipse, and for simplicity we assume that they are aligned with each other and with the angular momentum of binary C. We shall use Kepler angle-action variables; for binary A these are 
\begin{equation}
\lambda_A,\quad  \Lambda_A=\mu_A(GM_Aa_A)^{1/2}, \quad q_A=-\varpi_A,\quad Q_A=\Lambda_A[1-(1-e_A^2)^{1/2}].
\end{equation}
The Kepler Hamiltonian is
\begin{equation}
    H_{KA}\equiv -\frac{Gm_1m_2}{2a_A}=-\frac{G^2\mu_A^3M_A^2}{2\Lambda_A^2}.
    \label{eq:kepler}
\end{equation}
There are analogous expressions for binaries B and C. 

Our notation is mostly consistent with \cite{ham15} and \cite{bv18}. 

\medskip

\subsection{Dynamical stability}

The dynamical stability of hierarchical 2+2 quadruple systems has not been studied in detail. However, a reasonable approximation is to treat the quadruple as a pair of fictitious triple systems, in which first one of the close binaries and then the other is coalesced into a single star at their center of mass. Then the quadruple system is likely to be stable if both of the fictitious triple systems are stable. 

The stability criterion for triples depends strongly on the mutual inclination $I$ between the inner and outer binaries, since systems with $I\sim 90^\circ$ are subject to Lidov--Kozai oscillations that can excite the eccentricity of the inner binary. \cite{pet18} write the stability criterion in terms of the ratio of the periapsis of the outer binary to the apoapsis of the inner binary,
\begin{equation}
R_A\equiv \frac{a_C(1-e_C)}{a_A(1+e_A)};
\end{equation}
here we assume that the two members of binary B have been coalesced so we are testing the stability of binary A. 
Then, following \cite{ma01}, they argue that the system is usually stable if
\begin{equation}
R_A > \frac{2.8}{1+e_A}\left[\left(1+\frac{m_B}{m_A}\right) \frac{1+e_C}{(1-e_C)^{1/2}}\right]^{0.4}(1-0.3I/180^\circ).
\label{eq:ma}
\end{equation}
A similar stability criterion describes binary B. 

For small eccentricities, $\min(R_A,R_B)\sim \epsilon^{-1}$, so for the median $\epsilon\simeq 0.01$ in the Zasche et al.\ sample $R_A$ and $R_B$ are at least 100. Therefore most of the quadruple systems in this catalog are dynamically stable by a large margin. 

\medskip

\subsection{The Hamiltonian}

\smallskip

The resonant Hamiltonian in a 2+2 quadruple system is derived in Appendix \ref{sec:ham} and given by equation (\ref{eq:resham}): 
\begin{align}
  H_\mathrm{bb}=-\frac{3G\mu_A\mu_B}{32}\frac{a_A^2a_B^2}{r_C^5}\Big[& 3\cos(2\lambda_A-2\lambda_B) -9e_A\cos(\lambda_A-2\lambda_B+\varpi_A) \nonumber \\
 &\quad  - 10e_A\cos(\lambda_A-2\lambda_B -\varpi_A +2\phi_C) 
  + 3e_B\cos(3\lambda_B-2\lambda_A-\varpi_B)\Big],
\end{align}
with $r_C$ and $\phi_C$ the polar coordinates of binary C.  Here we have chosen the binary labels A and B so that the mean motion of A is larger than the mean motion of B, $n_A=\dot\lambda_A \ge n_B=\dot\lambda_B \gg n_C=\dot\lambda_C$. We have also kept only terms in the Hamiltonian that are zero or first order in the eccentricities $e_A$ and $e_B$; the justification is that (i) at the typical separation of the binaries in this sample, tidal friction rapidly damps the eccentricities (see \S\ref{sec:tidal}); (ii) we shall find that resonant capture is only efficient if at least one binary has near-zero eccentricity. The resonant Hamiltonian is smaller than the Kepler Hamiltonians for the two compact binaries by $\mbox{O}(\epsilon^5)$ where $\epsilon$ is defined by equation (\ref{eq:epsdef}), and we have dropped terms in the Hamiltonian that are higher order in $\epsilon$.

In terms of the action-angle variables for binaries A and B, the sum of the resonant and Kepler Hamiltonians is 
\begin{align}
  H&=H_{KA}+H_{KB}+H_\mathrm{bb}=-\frac{G^2\mu_A^3M_A^2}{2\Lambda_A^2}
  -\frac{G^2\mu_B^3M_B^2}{2\Lambda_B^2}\nonumber \\ &\quad-\frac{3\Lambda_A^4\Lambda_B^4}{32 G^3M_A^2M_B^2\mu_A^3\mu_B^3 r_C^5} \Big[ 3\cos(2\lambda_A-2\lambda_B)  -9(2Q_A/\Lambda_A)^{1/2}\cos(\lambda_A-2\lambda_B-q_{A}) \nonumber \\
 &\qquad\qquad - 10(2Q_A/\Lambda_A)^{1/2}\cos(\lambda_A-2\lambda_B +q_A +2\phi_C) 
  + 3(2Q_B/\Lambda_B)^{1/2}\cos(3\lambda_B-2\lambda_A+q_B)\Big].
  \label{eq:resaa}
\end{align}
Here we have approximated $Q_{A}$ as $\tfrac{1}{2}\Lambda_{A}e_A^2$, consistent with our expansion of the resonant Hamiltonian to first order in $e_A$. We make the same approximation for $Q_B$. 

Equation (\ref{eq:resaa}) for the resonant Hamiltonian contains one term independent of $Q$, which we call the ``corotation'' term, and three terms $\propto Q^{1/2}$, which we call ``Lindblad'' terms. There is one 3:2 Lindblad resonance and two 2:1 Lindblad resonances.

If binary C is on a circular orbit, then $\phi_C$ increases uniformly with time and we can choose the origin of time or azimuth such that $\phi_C=n_Ct$; also, the radius $r_C$ is constant and equal to the semimajor axis $a_C$. If C is eccentric, each resonant term is split into a multiplet consisting of a primary term independent of $e_C$ and secondary terms with strength proportional to $e_C$, $e_C^2$, etc., and frequencies differing by multiples of $n_C$. The resonant terms are split as
\begin{align}
\frac{A}{r_C^5}\cos(\Psi+p\phi_C) &\Longrightarrow 
\frac{A}{a_C^5}\Big\{\cos(\Psi+p\lambda_C) + e_C\big\{(\tfrac{5}{2}+p)\cos[\Psi+(p+1)\lambda_C+q_C]+ (\tfrac{5}{2}-p)\cos[\Psi+(p-1)\lambda_C-q_C]\big\} \nonumber \\
                       &\qquad\qquad + e_C^2\big\{(5-p^2)\cos(\Psi+p\lambda_C) + (5+\tfrac{25}{8}p+\tfrac{1}{2}p^2)\cos\big[\Psi+(p+2)\lambda_C+2q_C\big] \nonumber \\& \qquad\qquad + (5-\tfrac{25}{8}p+\tfrac{1}{2}p^2)\cos\big[\Psi+(p-2)\lambda_C-2q_C\big]\big\}\Big\} + \mbox{O}(e_C^3).
  \label{eq:split}
\end{align}
Here $p$ is any integer and $\Psi$ is any argument independent of the orbital phase of binary C. We may choose the origin of time or azimuth so the mean longitude of binary C is $\lambda_C=n_Ct$.

In summary, the possible resonances and their labels are:
\begin{align}
  \mbox{Corotation resonances:} \quad & 0=2n_A-2n_B+kn_C, \nonumber \\
  \mbox{3:2 Lindblad resonances:} \quad & 0=3n_B-2n_A-\dot\varpi_B+kn_C, \nonumber \\
  \mbox{2:1 Lindblad resonances (Type I):}\quad  & 0=2n_B-n_A-\dot\varpi_A+kn_C,\nonumber \\
 \mbox{2:1 Lindblad resonances (Type II):} \quad & 0=2n_B-n_A+\dot\varpi_A+(k-2)n_C.
  \label{eq:res}
\end{align}
Here $n_A\ge n_B$, $k$ is an integer, and the strength of the resonance is proportional to $e_C^{|k|}$.  All other resonances are $\mbox{O}(\epsilon^6)$ or higher. In principle $k$ can be any integer but in practice it is most likely that the system is trapped in one of the strongest resonances so $|k|$ should be small or zero, and much but not all of our discussion will be limited to the case $k=0$. The labels ``Type I'' and ``Type II'' are introduced for convenience in the discussions in \S \ref{sec:twoone}.

\medskip

\subsection{Tidal friction}

\smallskip

\label{sec:tidal}

Isolated close binary stars dissipate energy at fixed angular momentum, and evolve towards a minimum-energy state in which the orbit is circular and the spin angular velocities of both stars are the same as the orbital angular velocity (see \citealt{zahn2013} for a review).  The tidal evolution rate is a strong function of semimajor axis: in the simplest models of tidal friction, in which dissipation induces a fixed time lag to the tidal response \citep[e.g.,][]{hut81}, the characteristic evolution timescales for the semimajor axis $a$ and eccentricity $e$ are $\sim a^{8}$. The evolution is most rapid in stars with an outer convective zone, such as solar-type stars. 

In most cases the spin angular velocity of the stars evolves much faster than the semimajor axis or eccentricity, so the stars are found in a state of `pseudo-synchronization' in which the orbit-averaged tidal torque on their spin angular momenta vanishes. 

Tidal evolution timescales can be calibrated by determining the eccentricity as a function of orbital period for binaries in clusters and stellar populations with a variety of ages. Solar-type binaries with periods shorter than 7--8 days have circular orbits in clusters of all ages, probably because of tidal friction in the pre-main-sequence stage of stellar evolution. Halo binaries have circular orbits if the period is less than about 20 days \citep{mm04}. Presumably the orbits with periods between 8 and 20 days were circularized during the main-sequence phase that occupies most of the halo binaries' current lifetime of $\sim 10\Gyr$.

The eclipsing binaries in the sample of \cite{zasche19} have a median period of $1.90\dy$ with an interquartile range of 1.15--$4.1\dy$. Thus must of them must have synchronized and circularized in much less than $10\Gyr$. We shall find below that resonant capture can only occur from orbits with near-zero eccentricity; however, during evolution towards the synchronized, circularized state the eccentricity and semi-major axes evolve at similar rates \citep[e.g.,][]{hut81} so by the time the eccentricity is small enough for capture to occur the semimajor axis evolution has ceased. Therefore resonant capture requires a slow dissipative process that leads to continued evolution of the semimajor axis after the binary has achieved a synchronized circular state. There are at least two such processes: magnetic braking through stellar winds, and emission of gravitational radiation. 

\paragraph*{Magnetic braking:} For simplicity we model the binary as two identical solar-type stars. One star loses angular momentum through a wind. If the envelope of the star is convective -- for main-sequence stars this requires an effective temperature $<6250\mbox{\,K}$, corresponding to a spectral type of F8 and a mass of about $1.3\msun$ \citep{winn10} -- the wind is magnetically attached to the envelope of the rotating star, and angular-momentum loss through the wind is much more efficient than mass loss \citep{egg06}. The angular speed of isolated stars with convective envelopes decays as $\Omega(t)\propto t^{-1/2}$ (the Skumanich law). For a binary composed of solar-type stars, we can write $\Omega(t)=\Omega_\odot(t_\odot/t)^{1/2}$ where $\Omega_\odot$ and $t_\odot$ are the current angular speed and age of the Sun. Therefore
\begin{equation}
  \frac{d\Omega}{dt}= -\frac{\Omega_\odot t_\odot^{1/2}}{2t^{3/2}}=-\frac{\Omega^3}{2\Omega_\odot^2 t_\odot}.
  \end{equation}
The last equation expresses $\dot\Omega$ in terms of $\Omega$ because the strength of the wind is determined directly by the rotation rate of the star, and only indirectly by the age. The corresponding rate of angular-momentum loss is \citep{vb81}
\begin{equation}
  \frac{dL}{dt}=M_\odot k^2R_\odot^2\frac{d\Omega}{dt}=-\frac{M_\odot k^2R_\odot^2\Omega^3}{2\Omega_\odot^2 t_\odot},
\end{equation}
where $M_\odot$ is the mass and $kR_\odot$ is the radius of gyration of a solar-type star.

In a close binary the stars are maintained in synchronous rotation, so $\Omega=(2GM_\odot/a^3)^{1/2}$ where $a$ is the semimajor axis. The total angular momentum of the system is usually dominated by its orbital angular momentum, so $L=\half M_\odot(2GM_\odot a)^{1/2}$. Conservation of angular momentum then requires that
\begin{equation}
\frac{da}{dt}=-\frac{4k^2R_\odot^2 GM_\odot}{\Omega_\odot^2 t_\odot a^4}.
\end{equation}
This equation is easily solved to give \comment{2.9705}
\begin{equation}
  a=a_0(1-t/\tau_\mathrm{a,MB})^{1/5} \quad\mbox{where}\quad \tau_\mathrm{a,MB}=\frac{\Omega_\odot^2 t_\odot a_0^5}{20 k^2R_\odot^2GM_\odot}=3.0\Gyr\,\left(\frac{P_0}{2\dy}\right)^{10/3}.
\label{eq:mb}
\end{equation}
Here $P_0$ is the orbital period corresponding to the initial semimajor axis $a_0$ and we have used $k=0.26$,  $\Omega_\odot=2.87\times10^{-6}\mbox{\,s}^{-1}$, and $t_\odot=4.55\Gyr$. We conclude that solar-type binaries with periods $\lesssim 3\dy$ can experience substantial orbital decay due to stellar winds. This process is likely to be much less effective for stars earlier than F8, which have radiative envelopes. 

There are many oversimplifications in this analysis, including the restriction to identical solar-type stars and the assumption that only one star is emitting a wind.

\paragraph*{Gravitational radiation:} Two stars of masses $m_1$ and $m_2$ on a circular orbit of semimajor axis $a$ experience orbital decay due to gravitational radiation at a rate \citep{peters64}
\begin{equation}
  \frac{da}{dt}=-\frac{64 G^3m_1m_2(m_1+m_2)}{5c^5a^3}.
\end{equation}
For simplicity we assume that the two stars have the same mass $m$. This equation can be integrated to give
\begin{equation}
  a=a_0(1-t/\tau_\mathrm{a,GR})^{1/4} \quad\mbox{where}\quad \tau_\mathrm{a,GR}=\frac{5c^5a_0^4}{512 G^3 m^3}=3.77\times10^{11}\yr \left(\frac{\msun}{m}\right)^{5/3}\left(\frac{P}{2\dy}\right)^{8/3}.
\label{eq:gr}
\end{equation}
Thus solar-type binaries with periods $\lesssim 1\dy$ can experience substantial orbital decay due to gravitational radiation. 

\bigskip

\section{The corotation resonance}

\medskip

\label{sec:corot}

In terms of the orbital periods, the corotation resonance condition (\ref{eq:res}) can be written
\begin{equation}
  1-\frac{P_B}{P_A}=\frac{kP_B}{2P_C}.
  \label{eq:corot}
\end{equation}
At corotation, $P_B/P_C=2^{1/2}\epsilon^{3/2}$ where $\epsilon$ is defined in equation (\ref{eq:epsdef}).  Therefore the range of mean motions arising from resonance splitting is
\begin{equation}
  \left|1-\frac{P_B}{P_A}\right|= 0.7|k|\epsilon^{3/2}.
  \label{eq:split1}
\end{equation}

The orbital periods may also differ from the resonance condition (\ref{eq:corot}) because the libration amplitude is non-zero. To estimate the size of this effect, assume for simplicity that all four stars have the same mass. Then $a_A=a_B$ at corotation, and the maximum difference in mean motions is given by equation (\ref{eq:dnmax}) as 
\begin{equation}
  |n_A-n_B|_\mathrm{max}=n_B\frac{3^{3/2}}{4}\epsilon^{5/2},\quad\mbox{or}\quad \left|1-\frac{P_B}{P_A}\right|\le 1.30\,\epsilon^{5/2}.
   \label{eq:split2}
\end{equation}

The range of mean motions arising from resonance splitting (eq.\ \ref{eq:split1}) is larger than the range due to libration (eq.\ \ref{eq:split2}), by $\mbox{O}(\epsilon^{-1})\gg 1$. Therefore the maximum value of $|1-P_B/P_A|$ in the corotation resonance should be given by (\ref{eq:split1}). For the quadruple systems described after equation (\ref{eq:epsdefa}), the median value of $\epsilon$ is about 0.01 so we expect that $|1-P_B/P_A|$ should typically be less than $0.001|k|$. In the sample of \cite{zasche19} the two smallest values are 0.0081 and 0.0126 (the second of these is KIC 4247791, described in the Introduction). Thus it is unlikely that any of the quadruple systems in this sample is in the corotation resonance unless they are trapped in a resonance with $|k|\gg 1$ and $\epsilon$ is unusually large (eq.\ \ref{eq:ma} implies that $\epsilon \lesssim 0.27$ is required for dynamical stability, even if the masses are equal and all the orbits are circular). 

The absence of systems in the corotation resonance is consistent with the calculation in Appendix \ref{app:corot}, which shows that the probability of capture into this resonance is small.

Given that there is no observational or theoretical evidence for capture into the corotation resonance, we must attempt to explain why the distribution of period ratios in Figure \ref{fig:one} exhibits a peak near period ratios of unity. \cite{zasche19} argue that a peak in the period-ratio  distribution is expected even if the orbital periods in binaries A and B are uncorrelated. We have tested this hypothesis by randomly selecting pairs of binaries from the Zasche et al.\ catalog to create randomized quadruple systems. The result is shown in Figure \ref{fig:one} and although there \emph{is} a peak at 1:1 in the distribution of period ratios in the randomized catalog, the peak is much less prominent than observed. We have also compared the observed distribution to the distribution obtained by randomly selecting pairs of periods from a uniform distribution, $dn\propto dP$, over a range $P_\mathrm{min}\to P_\mathrm{max}$. In the limit where $P_\mathrm{max}/P_\mathrm{min}\gg1$ the distribution of period ratios $r=P_B/P_A$ should be $dn\propto r^{-2}$. We plot this distribution as a dotted line in Figure \ref{fig:one} and although the peak near $r=1$ is sharper it is still not as sharp as observed. 

Resonant capture could occur from weaker Lindblad-type resonance terms of the form $e_Ae_B\cos(2\lambda_A-2\lambda_B+\varpi_A-\varpi_B)$ but these are important only if one of the binaries has significant orbital eccentricity despite tidal friction. There are analogous inclination-type resonances if the orbital planes of binaries A, B, and C are not aligned. We also comment that (i) there is an obvious selection effect against finding systems in corotation resonance with small libration amplitude, since in this case the eclipses of binaries A and B will be superimposed; (ii) the catalog of \cite{zasche19} was compiled by manual inspection of over 26,000 eclipsing binaries in the OGLE database, so the selection effects cannot be described exactly.

Until we have a larger database of quadruple systems with well-understood selection effects, we cannot be certain whether the excess in systems near the corotation resonance is real, or what might be its explanation. 

\bigskip

\section{The 3:2 Lindblad resonance}

\medskip

\label{sec:threetwo}

The 3:2 Lindblad resonance conditions (\ref{eq:res}) can be written
\begin{equation}
 1-\frac{2P_B}{3P_A}=\frac{\dot\varpi_B P_B}{6\upi}-\frac{kP_B}{3P_C}.
\label{eq:split5}
\end{equation}
The range of mean motions due to resonance splitting is
\begin{equation}
  \left|1-\frac{2P_B}{3P_A}\right|=0.47|k|\epsilon^{3/2},
\label{eq:split3}
\end{equation}
assuming that all four stars have the same mass.  

We must also account for the precession rate $\dot\varpi$. The free precession rate is given by equation (\ref{eq:precrate}); if all the stars have equal mass and eccentricities near zero, $\dot\varpi_BP_B/(6\upi)=2^{-3/2}\epsilon^3$, which is negligible compared to the splitting (\ref{eq:split3}) when $\epsilon \ll1$. There is also a forced precession rate due to the resonant terms in the Hamiltonian. To obtain a rough estimate of the forced precession, we convert the resonant Hamiltonian to the dimensionless form (\ref{eq:second}). Then when the dimensionless squared eccentricity $R$ is of order unity, the precession rate $\dot\varpi \sim \epsilon_1\sim \epsilon^{10/3}$, which is also negligible compared to the splitting.  

As described in the preceding section, the median value of $\epsilon$ in the Zasche et al.\ catalog of doubly eclipsing binaries is about 0.01, so we expect $|1-2P_B/3P_A|\sim 0.0005|k|$ from resonance splitting, with negligible contributions from free or forced precession. In the sample of Zasche et al.\  the two smallest values are 0.0008 and 0.00005 (CzeV343 and BI 108, both described in the Introduction); these two systems are therefore likely to be in resonance. However, the peak at the 3:2 resonance in Figure \ref{fig:one} contains $\sim 8$ systems, which suggests that a few systems with larger values of $|1-2P_B/3P_A|$ must be trapped in resonance as well.  

As described in Appendix \ref{app:lind32}, capture into resonance requires that the parameter $\delta$ in the dimensionless Hamiltonian (\ref{eq:second}) increases with time, which in turn requires that $\dot n_A < \frac{3}{2}\dot n_B$ or $\dot n_A/n_A < \dot n_B/n_B$. Magnetic braking leads to $\dot n>0$ (loss of angular momentum makes the orbital frequency grow) so capture requires that the semimajor axis evolution time $\tau_a=|d\log a/dt|^{-1}$ in binary B is shorter than in A. For identical stars this is implausible, since evolution is generally faster in more compact binaries; however, given the strong dependence of the evolution rate -- whether by magnetic braking or gravitational radiation -- on stellar mass and other properties there should be many systems in which this condition is satisfied. 

Capture is certain if the evolution is slow enough and the initial value of the slow action or eccentricity of binary B is small enough. In the equal-mass case capture requires that the eccentricity $e_{B0}$ as the binary approaches resonance satisfies \comment{coefficient is 0.2329} 
\begin{equation}
    e_{B0} \lesssim \epsilon^{5/3}.
\label{eq:ecap}
\end{equation}
For $\epsilon=0.01$ this requires $e_{B0} \lesssim 0.0005$. Thus binary B must be on a nearly  circular orbit. 
    
Capture also requires that $0<d\delta/dt'=\epsilon_1^{-1}d\delta/dt <f_\tau$ where $f_\tau\simeq 2.75$ for nearly circular orbits based on numerical experiments (see also \citealt{quillen06} and \citealt{fried01}); thus in the equal-mass case we require
\begin{equation}
  0 < \frac{d \Delta}{dt} \lesssim \,n_B \epsilon^{20/3}\quad\mbox{where}\quad \Delta\equiv \frac{3n_B}{2n_A}-1.
  \label{eq:tidaltime}
\end{equation}
For example, if $\epsilon=0.01$ and the period of binary B is 3 days, the constraint is satisfied so long as the tidal evolution timescale $(d\Delta/dt)^{-1}$ exceeds a few times $10^{10}\yr$. Tidal evolution times longer than the age of the binary make capture unlikely, since the system must begin close to resonance in order to reach it by its present time. Therefore capture favors systems with $\epsilon$ somewhat larger than 0.01: for example, if $\epsilon=0.03$ the required tidal evolution timescale is longer than $\sim 20\Myr$.

All of these constraints on capture ignore eccentricity damping due to tidal friction. The characteristic eccentricity damping time $\tau_e$ can be much faster than the semimajor axis evolution time $\tau_a$ due to magnetic braking or gravitational radiation. The effect of eccentricity damping on Lindblad resonances is discussed by \cite{gs14}. Given that the strength of the resonance is of order $\epsilon^5$, they find that permanent trapping in resonance requires $\tau_e\lesssim \tau_a\epsilon^{10/3}$ or $\tau_e \lesssim 2\times10^3\yr(\tau_a/10^{10}\yr)(\epsilon/0.01)^{10/3}$. 

\bigskip

\section{The 2:1 Lindblad resonances}

\label{sec:twoone}

\medskip

The Type I 2:1 Lindblad resonance condition (\ref{eq:res}) can be written
\begin{equation}
 1-\frac{P_B}{2P_A}=\frac{\dot\varpi_A P_B}{4\upi}-\frac{kP_B}{2P_C}.
\end{equation}
The conditions for Type II resonance are
\begin{equation}
 1-\frac{P_B}{2P_A}=-\frac{\dot\varpi_A P_B}{4\upi}-\frac{(k-2)P_B}{2P_C}.
\end{equation}
The discussion in \S\ref{sec:threetwo} shows that variations in the resonance condition due to libration or apsidal precession are negligible, and resonance splitting leads to changes 
\begin{equation}
  \left|1-\frac{P_B}{2P_A}\right|\le-0.71\,\epsilon^{3/2}\left\{\begin{array}{ll}|k| &\mbox{Type I}, \\ |k+2| & \mbox{Type II}.\end{array}\right.
\label{eq:split33}
\end{equation}
  For the median value $\epsilon \simeq 0.01$, $|1-P_B/2P_A|$ should be less than $0.001|k|$ or $0.001|k+2|$. For comparison, the smallest value observed in the Zasche et al.\ catalog of doubly eclipsing binaries is about 0.014.  Thus it is unlikely that any of the quadruple systems in this sample is in a 2:1 Lindblad resonance.

  Capture into resonance requires that the parameter $\delta$ in the dimensionless Hamiltonian (\ref{eq:second}) increases with time. For the Type I resonance this requires that (eq.\ \ref{eq:delta21}) $\dot n_A < 2\dot n_B$, while for Type II resonances we need the opposite, $\dot n_A > 2\dot n_B$. For a quadruple system composed of identical stars, and assuming evolution leads to orbital decay, then tidal evolution would be in the direction required for Type II resonance capture, i.e., the semimajor axis evolution time of the larger binary is larger, $\tau_{a,B}>\tau_{a,A}$.  However, if the stars are not identical we may have $\tau_{a,B}<\tau_{a,A}$, in which case capture into the Type I resonance would occur.
  
The 2:1 Lindblad resonances differ from the 3:2 Lindblad resonances because the resonant arguments involve the longitude of periapsis of binary A rather than binary B. Therefore capture requires that the eccentricity of binary A rather than binary B is low enough.  Capture is certain if the evolution is slow enough and the initial value of the slow action is small enough; in the equal-mass case this requires $e_{A0}\lesssim \epsilon^{5/3}$ (cf.\ eq.\ \ref{eq:ecap}). Thus binary A must be on a circular orbit. 
    
Capture also requires that $0<d\delta/dt'=\epsilon_1^{-1}d\delta/dt <f_\tau$ where $f_\tau=2.75$ for nearly circular orbits; thus in the equal-mass case we require 
\begin{align}
  0 > \frac{d \Delta}{dt} &\gtrsim -n_B \epsilon^{20/3}\quad&& \mbox{Type I}, \nonumber \\
  0 < \frac{d \Delta}{dt} & \lesssim n_B \epsilon^{20/3}\quad&& \mbox{Type II} .
  \end{align}
where $\Delta=n_A/2n_B-1$. The constraints on the tidal evolution time are similar to those discussed after equation (\ref{eq:tidaltime}).

\bigskip

\section{Discussion}

\medskip

\label{sec:disc}

We have shown that resonant capture can occur in 2+2 quadruple stellar systems. The strength of the resonant terms in the Hamiltonian can be parametrized by $\epsilon$, the ratio of the semimajor axes of the compact binaries A and B to the semimajor axis of the outer binary C (eq.\ \ref{eq:epsdef}). The dimensionless strength of the resonant Hamiltonian is of order $\epsilon^5$ (hexadecapole) which is typically quite weak: $\epsilon\lesssim 0.3$ is required for stability of a hierarchical quadrupole system (eq.\ \ref{eq:ma}) and the median value in the sample of \citet[][Table 4]{zasche19} is much smaller, $\epsilon\sim 0.01$. In these circumstances permanent capture can only occur with significant probability at the 3:2 or 2:1 Lindblad resonance (eq.\ \ref{eq:res}) and then only if (i) the tidal evolution is in the correct direction, such that $d\delta/dt>0$ (eqs.\ \ref{eq:xxx} and \ref{eq:delta21}); (ii) at least one of the binaries is on a nearly circular orbit, $e\lesssim \epsilon^{5/3}$, (iii) the semimajor axis evolution time $\tau_a$ is shorter than the age of the stars but longer than $\sim P\epsilon^{-20/3}$ where $P$ is the orbital period of an inner binary; (iv) the eccentricity damping time $\tau_e\lesssim \tau_a\epsilon^{10/3}$. 

Most known 2+2 resonant quadruple systems are doubly eclipsing binaries. These have orbital periods short enough that they should have synchronized and circularized rapidly (\S\ref{sec:tidal}). Thus condition (ii), a nearly circular orbit, is not difficult to achieve. Once the orbits are circularized, further semimajor axis evolution can occur through magnetic braking or gravitational radiation. As outlined in \S\ref{sec:tidal}, magnetic braking generally leads to stronger evolution than gravitational radiation, at least for low-mass stars with convective envelopes, and provides a plausible mechanism to satisfy condition (iii), at least in some cases. Condition (i) is satisfied for a quadruple system of identical stars only for Type II 2:1 Lindblad resonances, but since the semimajor axis evolution is a strong function of stellar mass and other parameters there should be many systems in which the condition is satisfied for any Lindblad resonance. 

The comparison of our models for resonant capture with the small sample of known resonant 2+2 systems in \cite{zasche19} raises a number of unanswered issues:

\begin{itemize}

\item There is an enhanced density of systems \emph{near} the 1:1 resonance (see Figure \ref{fig:one}), but none of these systems are found \emph{in} the resonance. The absence of systems in the 1:1 resonance is not surprising, since capture into this resonance is likely to be extremely inefficient, but we do not have an explanation for the concentration of systems nearby. 

\item Only 2 of the $\sim 8$ quadruple systems in the peak near the 3:2 resonance appear to be close enough to exact commensurability to be in the resonance. 

\item There appears to be a gap, rather than an enhancement, in the density of systems near the 2:1 Lindblad resonance. 

\item Magnetic braking should be much less efficient in main-sequence stars earlier than about F8, which do not have convective envelopes, but systems like V994 Her which contain four stars no later than A0 are found close to the 3:2 resonance. 

\end{itemize}

The situation is reminiscent of the distribution of period ratios in multi-planet systems discovered by the {\it Kepler} spacecraft, which also exhibits a sharp peak near the 3:2 resonance. The peak is displaced slightly outside the resonance (i.e., period ratios larger than 1.5). There is also a gap just inside the 2:1 resonance and a peak just outside, and weaker evidence for peaks outside other resonances \citep{fab14}.  The physical mechanism leading to resonance capture in planetary systems is probably disk migration, which differs from tidal friction in several respects --in particular the evolution time is much shorter.

We now briefly discuss the example systems described in the Introduction. The most complete information is available for V994 Her. Given the masses and orbital periods from \cite{zasche16} the ratio of semimajor axes is $\epsilon=a_B/a_c=0.014$. Resonant capture requires that the initial eccentricity $e_{B0}\lesssim \epsilon^{5/3}\simeq 0.001$ (eq.\ \ref{eq:ecap}); this is much smaller than the observed eccentricity $e_B=0.03$ but perhaps the eccentricity has been pumped up after capture. A more serious concern is that the distance of the star from resonance, $1-2P_B/3P_A = 0.0220$, requires that capture has occurred in a harmonic with $k\simeq 30$ (eq.\ \ref{eq:split5}), which is improbable even though the eccentricity of the outer binary is large, $e_c= 0.76\pm 0.07$. Thus we argue that V994 Her is probably not in resonance. Similarly OGLE BLG-ECL-145467 is too far from the 3:2 resonance, and KIC 4247791 is too far from the 1:1 resonance -- and in addition, as we have seen, capture at the 1:1 resonance is highly improbable. In contrast CzeV343 and BI 108 have period ratios sufficiently close to 3:2 that they probably \emph{are} in resonance, although in this case the eccentricity $e_B=0.18$ must have been excited after resonant capture.

We have simplified our analysis in several important ways. We have assumed that the inner and outer binary orbits are coplanar, but all we know is that the normals to both inner binary orbits are nearly perpendicular to the line of sight, since they eclipse. In non-coplanar systems there are additional inclination resonances and novel phenomena such as Lidov--Kozai resonances are possible.
We have generally assumed that the orbit of the outer binary is circular, but according to the discussion around equation (\ref{eq:split}) a non-zero eccentricity only splits the resonances, and we have estimated the sizes of the splittings in equations (\ref{eq:split1}), (\ref{eq:split3}), and (\ref{eq:split33}) and found them to be small. We have also neglected resonant terms that are weaker than the dominant resonances by of order $\epsilon$, but the dominant resonances are already weak enough that their even weaker cousins are not likely to have a significant dynamical effect. 

We have presented a framework for studying resonances in 2+2 quadruple systems, but further progress will require large samples of doubly eclipsing binaries constructed with well-defined selection criteria. These samples should allow us to address questions such as: at what resonances are there peaks in the distribution of period ratios? How close are the period ratios to exact resonance? How does the likelihood of capture depend on the spectral type and other properties of the stars? The unique phenomenon of resonant capture in quadruple stars promises a new window into the formation and long-term evolution of single and binary stars. 

\section*{Acknowledgements}

I thank Todd Thompson introducing me to this problem and for many discussions, and S\-lawomir Breiter and David Vokrouhlick\'y for sending us their unpublished notes on this subject, and for thoughtful comments on the paper.

\appendix

\bigskip

\section{The Hamiltonian}

\medskip

\label{sec:ham}

The Hamiltonian for a hierarchical quadruple system is given in equation (A14) of \cite{ham15},
\begin{equation}
    H=H_{KA}+H_{KB}+H_{KC}+H_{Q,AC}+H_{Q,BC}+H_\mathrm{bb}.
\end{equation}
The first three terms are the Kepler Hamiltonian of equation (\ref{eq:kepler}) for binary A and its analogs for B and C. Since binary C has much more angular momentum than A or B, its orbit is only weakly affected by the evolution of A and B; therefore we can drop the term $H_{KC}$. The Hamiltonian $H_{Q,AC}$ arises from quadrupole and higher order interactions between binaries A and C, and depends only on $\bra$ and $\brc$. Since the orbital periods of binaries A and C are very different, $H_{Q,AC}$ has no resonant terms. Moreover it is smaller than the Kepler Hamiltonian $H_{KA}$ by $\mbox{O}(\epsilon^3)$. Its only significant effect is to induce a slow apsidal precession in binary A. To describe this precession the orbit-averaged Hamiltonian $\overline H_{Q,AC}$ is sufficient. Keeping only the quadrupole terms and restricting equation (A10b) of \cite{ham15} to the coplanar case, we find 
\begin{align}
  \overline H_{Q,AC}&=-\frac{Gm_1m_2(m_3+m_4)}{m_1+m_2}\frac{a_A^2}{8(1-e_C^2)^{3/2}a_C^3}(2+3e_A^2)\nonumber \\
&=-\frac{Gm_1m_2(m_3+m_4)}{m_1+m_2}\frac{a_A^2} {8(1-e_C^2)^{3/2}\Lambda_A^2a_C^3}(2\Lambda_A^2+6Q_A\Lambda_A-3Q_A^2) 
\end{align}
with an analogous expression for $\overline H_{Q,BC}$. The precession rate is
\begin{equation}
  \dot\varpi_A=-\dot q_A=-\frac{\p \overline H_{Q,AC}}{\p Q_A}=\frac{Gm_1m_2(m_3+m_4)}{m_1+m_2}\frac{3a_A^2} {4(1-e_C^2)^{3/2}\Lambda_A^2a_C^3}(\Lambda_A-Q_A).
  \label{eq:precrate}
  \end{equation}

The final term in the Hamiltonian arises from interactions between binary A and binary B, 
\begin{align}
  H_\mathrm{bb}&=-\frac{3G\mu_A\mu_B}{4\rc^7}\Big[\ra^2\rb^2\rc^2-5\rb^2(\bra\cdot\brc)^2
  -5\ra^2(\brb\cdot\brc)^2 \nonumber \\&\qquad +\frac{35}{\rc^2}(\bra\cdot\brc)^2(\brb\cdot\brc)^2+2\rc^2(\bra\cdot\brb)^2-20(\bra\cdot\brc)(\brb \cdot\brc)(\bra\cdot\brb)\Big].
\end{align}
This is proportional to $r_C^{-5}$ and thus is a hexadecapole term, smaller than the Kepler Hamiltonians by $\mbox{O}(\epsilon^5)$. 

We assume that the orbits are coplanar, with zero inclination, and expand to first order in the eccentricities of binaries A and B. We write the polar coordinates of binary C as $r_C, \phi_C$. After dropping a constant term, 
\begin{align}
  H_\mathrm{bb}&=-\frac{3G\mu_A\mu_B}{32}\frac{a_A^2a_B^2}{r_C^5}\Big\{
  3\cos(2\lambda_A-2\lambda_B) + 10\cos(2\lambda_A-2\phi_C)+ 10\cos(2\lambda_B-2\phi_C) + 35\cos(2\lambda_A+2\lambda_B-4\phi_C) \nonumber \\ &\quad +e_A[10\cos(3\lambda_A-\varpi_A-2\phi_C) -10\cos(\lambda_A+2\lambda_B-\varpi_A-2\phi_C) + 35\cos(3\lambda_A+2\lambda_B-\varpi_A-4\phi_C) \nonumber \\ & \qquad\quad + 3\cos(3\lambda_A-2\lambda_B-\varpi_A) -12\cos(\lambda_A-\varpi_A) -9\cos(\lambda_A-2\lambda_B+\varpi_A) - 30\cos(\lambda_A+\varpi_A-2\phi_C) \nonumber \\ &\qquad\quad - 10\cos(\lambda_A-2\lambda_B -\varpi_A +2\phi_C) - 105\cos(\lambda_A+2\lambda_B+\varpi_A-4\phi_C)] \nonumber \\
  &\quad + e_B[10\cos(3\lambda_B-\varpi_B-2\phi_C) - 10\cos(\lambda_B+2\lambda_A-\varpi_B-2\phi_C) + 35\cos(3\lambda_B+2\lambda_A-\varpi_B-4\phi_C)  \nonumber \\ & \qquad\quad + 3\cos(3\lambda_B-2\lambda_A-\varpi_B) -12\cos(\lambda_B-\varpi_B) -9\cos(\lambda_B-2\lambda_A+\varpi_B) - 30\cos(\lambda_B+\varpi_B-2\phi_C) \nonumber \\ &\qquad\quad - 10\cos(\lambda_B-2\lambda_A -\varpi_B +2\phi_C)- 105\cos(\lambda_B+2\lambda_A+\varpi_A-4\phi_C)]\Big\}
\end{align}

Now choose the labels so the mean motion of A is larger than the mean motion of B, $n_A=\dot\lambda_A \ge n_B=\dot\lambda_B \gg n_C=\dot\lambda_C$. Then eliminate all non-resonant terms, by which we mean all terms in which the rate of change of the argument of the cosine cannot vanish:
\begin{align}
  H_\mathrm{bb}=-\frac{3G\mu_A\mu_B}{32}\frac{a_A^2a_B^2}{r_C^5}\Big[& 3\cos(2\lambda_A-2\lambda_B) -9e_A\cos(\lambda_A-2\lambda_B+\varpi_A) \nonumber \\
 &\quad  - 10e_A\cos(\lambda_A-2\lambda_B -\varpi_A +2\phi_C) 
   + 3e_B\cos(3\lambda_B-2\lambda_A-\varpi_B)\Big].
   \label{eq:resham}
\end{align}

\bigskip

\section{Libration and capture in the corotation resonance}

\label{app:corot}

\medskip

The resonant Hamiltonian is given by the Kepler and corotation terms of equation (\ref{eq:resaa}): 
\begin{align}
  H&=-\frac{G^2\mu_A^3M_A^2}{2\Lambda_A^2} -\frac{G^2\mu_B^3M_B^2}{2\Lambda_B^2} -\frac{9\Lambda_A^4\Lambda_B^4}{32 G^3M_A^2M_B^2\mu_A^3\mu_B^3 a_C^5}\cos(2\lambda_A-2\lambda_B) . 
\end{align}
Here we have assumed that the orbit of binary C is circular; the results below can be generalized to non-circular orbits as outlined in the discussion surrounding equation (\ref{eq:split}).

We introduce a slow action and angle $(S,s)$ and a fast action and angle $(F,f)$. Using the canonical transformation defined by the generating function 
\begin{equation}
    \mathfrak{S}(S,F,\lambda_A,\lambda_B)=2S(\lambda_A-\lambda_B) + F\lambda_B,
\end{equation}
we have
\begin{equation}
\Lambda_A=\frac{\p \mathfrak{S}}{\p\lambda_A}=2S, \quad
\Lambda_B=\frac{\p \mathfrak{S}}{\p\lambda_B}=F-2S,\quad
s=\frac{\p \mathfrak{S}}{\p S}=2(\lambda_A-\lambda_B), \quad f=\frac{\p \mathfrak{S}}{\p F}=\lambda_B. 
\end{equation}
Thus
\begin{equation}
    H=-\frac{\alpha}{4S^2}-\frac{\beta}{(F-2S)^2}-\gamma\cos s,
\end{equation}
where $\alpha$ and $\beta$ are constants. The function $\gamma$ depends on the actions but this dependence is slow so for the moment we may treat it as a constant as well. The action $F$ is a constant of motion because the Hamiltonian is independent of its conjugate angle $f$. Let $S=S_0+\Delta S$ and expand to second order in $\Delta S$. Dropping an unimportant constant,
\begin{equation}
    H=\Delta S\left[\frac{\alpha}{2S_0^3}-\frac{4\beta}{(F-2S_0)^3}\right] -(\Delta S)^2\left[\frac{3\alpha}{4S_0^4}+\frac{12\beta}{(F-2S_0)^4}\right] -\gamma \cos s.
\end{equation}
We may choose $S_0$ to correspond to the resonance condition $n_A=n_B$. Then the coefficient of $\Delta S$ vanishes and we have 
\begin{align}
    H(\Delta S,s)&=-(\Delta S)^2\left(\frac{6}{\mu_Aa_A^2}+\frac{6}{\mu_Ba_B^2}\right) -\gamma \cos s\nonumber \\     &\equiv -\rho(\Delta S)^2 - \gamma\cos s
    \label{eq:coro}
\end{align}
where $\rho$ and $\gamma$ are both positive. The equations of motion are the pendulum equations,
\begin{equation}
  \frac{ds}{dt} = -2\rho\Delta S, \quad \frac{d\Delta S}{dt}=-\gamma\sin s.
  \label{eq:eqmotpend}
\end{equation}
  
Since $\rho$ and $\gamma$ are positive, there is an unstable equilibrium at $s=\Delta S=0$, corresponding to energy $E_\mathrm{sep}=H(0,0)=-\gamma$. Trajectories emerging from this unstable equilibrium define the separatrix, which separates librating from circulating motion. Librating trajectories have $E>E_\mathrm{sep}$ or $\rho(\Delta S)^2+\gamma(\cos s-1)<0$. The maximum value of $|\Delta S|$ in a librating trajectory occurs on the separatrix at $s=\upi$ and is given by $|\Delta S|_\mathrm{max}=(2\gamma/\rho)^{1/2}$. The corresponding maximum value of the difference in mean motions $n_A-n_B$ is
\begin{equation}
  |n_A-n_B|_\mathrm{max}=\half|\dot s|_\mathrm{max}=\rho|\Delta S|_\mathrm{max}=(2\gamma\rho)^{1/2}=\frac{3^{3/2}\mu_A^{1/2}\mu_B^{1/2}G^{1/2}a_Aa_B}{2^{3/2}a_C^{5/2}}\left(\frac{1}{\mu_Aa_A^2}+\frac{1}{\mu_Ba_B^2}\right)^{1/2}.
  \label{eq:dnmax}
\end{equation}

Now suppose that the slow action is subject to tidal forces causing it to evolve at a rate $\dot S_\mathrm{tid}(t)$ (for simplicity we ignore tidal evolution of the fast action; this is straightforward to include but does not produce a qualitative change in the results). Then the equations of motion are modified from (\ref{eq:eqmotpend}) to
\begin{equation}
     \frac{ds}{dt}  = -2\rho\Delta S, \quad \frac{d\Delta S}{dt} =-\gamma\sin s+\dot S_\mathrm{tid}(t).
\end{equation}
These equations of motion can be derived from a Hamiltonian
\begin{equation}
    H_\mathrm{tid}=-\rho[\widetilde S+ S_\mathrm{tid}(t)]^2 -\gamma\cos s,
\end{equation}
if $\widetilde S$ is conjugate to $s$ and we relate the trajectories by $\Delta S(t)=\widetilde S+S_\mathrm{tid}(t)$.

Now rescale the variables according to
\begin{equation}
    dt'=2\rho(t) dt,\quad I=-\epsilon \widetilde S, \quad  \phi=\upi+\epsilon s, \quad c(t')=-\epsilon S_\mathrm{tid}(t), \quad b(t')=\frac{\gamma(t)}{2\rho(t)}, \quad \epsilon=\mbox{sgn}\,(\dot S_\mathrm{tid}). 
    \label{eq:rescalec}
\end{equation}
The Hamiltonian in the rescaled time is
\begin{equation}
  H'(R,r,t')=\half[I+c(t')]^2-b(t')\cos \phi.
\end{equation}
Here $I$ and $\phi$ are the rescaled momentum and coordinate, and $b(t')>0$, $dc(t')/dt'<0$. 

The behavior of trajectories governed by this Hamiltonian is described by \cite{yoder79} and \cite{henrard82}. Henrard's balance-of-energy integrals are 
\begin{equation}
    B_1=2\upi\frac{d c}{dt'} -\frac{4}{b^{1/2}}\frac{db}{dt'}, \quad B_2=-2\upi\frac{d c}{dt'} -\frac{4}{b^{1/2}}\frac{db}{dt'}.
\end{equation}
Then assuming $B_1<0$ there are three possible outcomes as the system passes through resonance:
\begin{itemize}
    \item if $B_2<0$ the system is captured into resonance;
    \item if $B_2>0$ and $B_1+B_2>0$ the system is not captured;
    \item if  $B_2>0$ and $B_1+B_2<0$ the system is captured with probability $1+B_2/B_1$.
\end{itemize}

If the masses $m$ of all four stars are the same, $b=3\cdot 2^{-11} Gm^3a_A^6/a_C^5$ and we have
\begin{equation}
\frac{d c(t')}{dt'}\sim \frac{G^{1/2}m^{5/2}a_A^{5/2}}{\tau}, \quad \frac{1}{b^{1/2}}\frac{db}{dt'}\sim \frac{G^{1/2}m^{5/2}a_A^5}{a_C^{5/2}\tau},
\end{equation}
where $\tau$ is the tidal evolution time (in units of $t$, not $t'$). The ratio of the second term to the first is $\sim a_A^{5/2}/a_C^{5/2}=\epsilon^{5/2}\ll1$. Since $dc(t')/dt'<0$, we conclude that $B_1<0$, $B_2>0$, $|B_1+B_2|/|B_1|\sim \epsilon^{5/2}$. Whether or not capture can occur at the corotation resonance depends on the sign of $db(t')/dt'$, but even if capture occurs it does so with a very low probability, of order $\epsilon^{5/2}$.

\bigskip

\section{Libration and capture in the 3:2 Lindblad resonance}

\label{app:lind32}

\medskip

The Kepler and resonant terms in the Hamiltonian are 
\begin{align}
  H=-\frac{G^2\mu_A^3M_A^2}{2\Lambda_A^2} -\frac{G^2\mu_B^3M_B^2}{2\Lambda_B^2}
-\frac{9\Lambda_A^4\Lambda_B^4}{32 G^3M_A^2M_B^2\mu_A^3\mu_B^3 a_C^5}(2Q_B/\Lambda_B)^{1/2}\cos(3\lambda_B-2\lambda_A+q_B),
\end{align}
in which we have assumed that the orbit of binary C is circular.

Now introduce three new coordinate-momentum pairs $(s,S)$ (`S' for `slow'), $(f_A,F_A)$, $(f_B,F_B)$ (`F' for `fast'), defined by the canonical transformation
\begin{equation}
    \mathfrak{S}(S,F_A,F_B,\lambda_A,\lambda_B)=S(3\lambda_B-2\lambda_A+q_B)+F_A\lambda_A+F_B\lambda_B.
\end{equation}
Then 
\begin{equation}
\Lambda_A=\frac{\p \mathfrak{S}}{\p\lambda_A}=F_A-2S, \quad
\Lambda_B=\frac{\p \mathfrak{S}}{\p\lambda_B}=F_B+3S,\quad
Q_B=\frac{\p \mathfrak{S}}{\p q_B}=S,
\end{equation}
\begin{equation}
s=\frac{\p \mathfrak{S}}{\p S}=3\lambda_B-2\lambda_A+q_B, \quad  f_A=\frac{\p \mathfrak{S}}{\p F_A}=\lambda_A, \quad
f_B=\frac{\p \mathfrak{S}}{\p F_B}=\lambda_B. 
\end{equation}
Thus
\begin{equation}
    H=-\frac{\alpha}{(F_A-2S)^2}-\frac{\beta}{(F_B+3S)^2}-\gamma S^{1/2}\cos s
\end{equation}
where $\alpha$, $\beta$, and $\gamma$ are positive constants. Since the Hamiltonian is independent of the angles $f_{A,B}$, the actions $F_{A,B}$ are constant. Moreover since the eccentricities are assumed to be small, $S$ is also small so we can expand the Kepler part of the Hamiltonian to second order in $S$:
\begin{equation}
    H=\mbox{const}+S\left(\frac{6\beta}{F_B^3}-\frac{4\alpha}{F_A^3}\right) - S^2\left(\frac{12\alpha}{F_A^4}+\frac{27\beta}{F_B^4}\right) -\gamma S^{1/2}\cos s.
\end{equation}
Note that the coefficient of the term linear in $S$ is equal to $3n_B-2n_A$ when $S$ is small, so this term is zero as the system passes through resonance.

Now rescale the variables according to
\begin{equation}
  t'=\epsilon_1t,\quad R=\epsilon_2S, \quad r=\upi-s
\end{equation}
where 
\begin{equation}
\epsilon_1=\half\gamma^{2/3}\left(\frac{12\alpha}{F_A^4}+\frac{27\beta}{F_B^4}\right)^{1/3}\!\!, \quad \epsilon_2=\frac{2}{\gamma^{2/3}}\left(\frac{12\alpha}{F_A^4}+\frac{27\beta}{F_B^4}\right)^{2/3}\!\!, \quad 1+\delta =-\frac{2}{3\gamma^{2/3}}\left(\frac{4\alpha}{F_A^3}-\frac{6\beta}{F_B^3}\right)\left(\frac{12\alpha}{F_A^4}+\frac{27\beta}{F_B^4}\right)^{-1/3}.
\label{eq:eps32}
\end{equation}
The rescaled Hamiltonian is
\begin{equation}
  H'(R,r,t')=-3(1+\delta)R + R^2-2^{3/2}R^{1/2}\cos r.
\label{eq:second}
  \end{equation}.
In evaluating the fast actions, since $e$ is small we can replace $F_A$ by $\Lambda_A$ and $F_B$ by $\Lambda_B$. Then
\begin{equation}
    \frac{\alpha}{F_A^4}=\frac{1}{2a_A^2{\mu_A}}, \quad  \frac{\beta}{F_B^4}=\frac{1}{2a_B^2{\mu_B}}, \quad 1+\delta= \frac{2^{13/3}}{3^{10/3}}\left(1+\frac{4{\mu_B}a_B^2}{9{\mu_A}a_A^2}\right)^{-1/3}\frac{M_B^{1/6}M_A^{1/2}}{{\mu_A}^{2/3}}\frac{a_C^{10/3}}{a_A^{17/6}a_B^{1/2}}\frac{3n_B-2n_A}{n_A}.
\label{eq:xxx}
\end{equation}
In the simple case where all four stellar masses are equal, the last expression simplifies to 
\begin{equation}
1+\delta=\frac{4.645}{\epsilon^{10/3}}\left(1-\frac{2P_B}{3P_A}\right),
\label{eq:deldef}
\end{equation}
where $\epsilon$ is defined by equation (\ref{eq:epsdef}).

Equation (\ref{eq:second}) is the `second fundamental model of resonance' described by \citet{hl83}. The system passes through resonance when $1+\delta(t')=0$. \cite{hl83} show from the theory of adiabatic invariance that capture into resonance is certain if three conditions are satisfied:
\begin{itemize}
  
    \item $d\delta/dt'>0$, i.e., $\delta(t')$ increases through zero as the time $t$ or rescaled time $t'$ increases. If $\delta(t')$ decreases through zero, resonance capture cannot occur. Thus capture requires that $\dot n_A < \frac{3}{2}\dot n_B$.
    
    \item the initial action $R_0$ when the system is far from resonance is less than 3. For $R_0>3$ capture is probabilistic, and the capture probability declines with increasing $R_0$. An analytic expression for the capture probability is given by \cite{bg84}. Rescaling to physical variables, $R_0<3$ corresponds to
      \begin{equation}
        e_{B0}^2 < \frac{3^{1/3}}{2^{7/3}}\frac{\mu_A^{2/3}}{M_B^{2/3}}\frac{a_B^2 a_A^{4/3}}{a_C^{10/3}}\left(1+\frac{4\mu_Ba_B^2}{9\mu_Aa_A^2}\right)^{-2/3}.
        \end{equation}
    
    \item $\delta(t')$ varies slowly enough; in particular, when $d\delta/dt'\gg 1$ capture does not occur. When $R_0\ll1$, capture requires $0<d\delta/dt' < f_\tau$ where $f_\tau=2.75$ from numerical integrations. When $R_0$ is non-zero, capture is probabilistic when $d\delta/dt'$ is of order unity.
    
\end{itemize}

\bigskip

\section{Libration and capture in the 2:1 Lindblad resonances}

\medskip

\label{se:apptwoone}

There is a pair of resonant terms, for which the Hamiltonians can be written 
\begin{align}
  H=-\frac{G^2\mu_A^3M_A^2}{2\Lambda_A^2}
  -\frac{G^2\mu_B^3M_B^2}{2\Lambda_B^2}+\frac{k_\pm \Lambda_A^{7/2}\Lambda_B^4}{G^3M_A^2M_B^2\mu_A^3\mu_B^3a_C^5}Q_A^{1/2} \cos(\lambda_A-2\lambda_B\mp q_A+\omega_\pm t),
\end{align}
where $k_+=3^3/2^{9/2}$, $k_-=15/2^{7/2}$, $\omega_+=0$, $\omega_-=2n_C$, and the upper and lower signs refer to Type I and Type II resonances, respectively. 
Using the canonical transformation
\begin{equation}
    \mathfrak{S}(S,F_A,F_B,\lambda_A,\lambda_B)=S(\mp\lambda_A\pm2\lambda_B+q_A\mp\omega_\pm t)+F_A\lambda_A+F_B\lambda_B
\end{equation}
we have
\begin{equation}
\Lambda_A=\frac{\p \mathfrak{S}}{\p\lambda_A}=F_A\mp S, \quad
\Lambda_B=\frac{\p \mathfrak{S}}{\p\lambda_B}=F_B\pm 2S,\quad
Q_A=\frac{\p \mathfrak{S}}{\p q_B}=S,
\end{equation}
\begin{equation}
s=\frac{\p \mathfrak{S}}{\p S}=\mp\lambda_A\pm2\lambda_B+q_A\mp\omega_\pm t, \quad  f_A=\frac{\p \mathfrak{S}}{\p F_A}=\lambda_A, \quad
f_B=\frac{\p \mathfrak{S}}{\p F_B}=\lambda_B. 
\end{equation}
Thus
\begin{equation}
    H=-\frac{\alpha}{(F_A\mp S)^2}-\frac{\beta}{(F_B\pm 2S)^2}+\gamma_\pm S^{1/2}\cos s\mp\omega_\pm S,
\end{equation}
where $\alpha$, $\beta$, and $\gamma_\pm$ are constants. The actions $F_{A,B}$ are constant. Expanding to second order in $S$ and dropping an unimportant constant,
\begin{equation}
    H=S\left(\mp\frac{2\alpha}{F_A^3}\pm\frac{4\beta}{F_B^3}\mp\omega_\pm\right) - S^2\left(\frac{3\alpha}{F_A^4}+\frac{12\beta}{F_B^4}\right) +\gamma_\pm S^{1/2}\cos s.
\end{equation}

Now rescale the variables according to
\begin{equation}
  t'=\epsilon_1t,\quad R=\epsilon_2S, \quad r=-s
\end{equation}
where 
\begin{equation}
\epsilon_1=\half\gamma_\pm^{2/3}\left(\frac{3\alpha}{F_A^4}+\frac{12\beta}{F_B^4}\right)^{1/3}\!\!, \quad \epsilon_2=\frac{2}{\gamma_\pm^{2/3}}\left(\frac{3\alpha}{F_A^4}+\frac{12\beta}{F_B^4}\right)^{2/3}\!\!, \quad 1+\delta =\mp\frac{2}{3\gamma_\pm^{2/3}}\left(\frac{2\alpha}{F_A^3}-\frac{4\beta}{F_B^3}+\omega_\pm\right)\left(\frac{3\alpha}{F_A^4}+\frac{12\beta}{F_B^4}\right)^{-1/3}\!\!.
\end{equation}
The rescaled Hamiltonian is given by equation (\ref{eq:second}). In evaluating the fast actions, since $e$ is small we can replace $F_A$ by $\Lambda_A$ and $F_B$ by $\Lambda_B$. Then
\begin{equation}
  \frac{\alpha}{F_A^4}=\frac{1}{2a_A^2{\mu_A}}, \quad  \frac{\beta}{F_B^4}=\frac{1}{2a_B^2{\mu_B}}, \quad 1+\delta= \mp \frac{2^{2/3}}{3^{4/3}k_\pm^{2/3}}\left(1+\frac{{\mu_B} a_B^2}{4{\mu_A} a_A^2}\right)^{-1/3}\frac{M_A^{1/6}M_B^{1/2}}{{\mu_A}^{1/3}{\mu_B}^{1/3}}\frac{a_C^{10/3}}{a_A^{7/6}a_B^{13/6}}\frac{n_A-2n_B+\omega_\pm}{n_B}.
  \label{eq:delta21}
    \end{equation}

First consider the Type I resonance (upper sign). Capture into resonance requires that $\delta$ is increasing so we need $\dot n_A < 2\dot n_B$. Capture is certain if 
\begin{equation}
    e_{A0}^2 < \frac{3^{7/3}}{2^{11/3}}\frac{{\mu_B}^{4/3}}{M_A^{2/3}{\mu_A}^{2/3}}\frac{a_A^{2/3}a_B^{8/3}}{a_C^{10/3}}\left(1+\frac{{\mu_B} a_B^2}{4{\mu_A} a_A^2}\right)^{-2/3},
\label{eq:ecapi}
  \end{equation}
 which requires that the orbit of binary A is nearly circular. 

Next consider the Type II resonance. Capture into resonance requires that $\delta$ is increasing so we need $\dot n_A > 2\dot n_B$. Capture is certain if 
\begin{equation}
   e_{A0}^2 < \frac{3\cdot 5^{2/3}}{8}\frac{{\mu_B}^{4/3}}{M_A^{2/3}{\mu_A}^{2/3}}\frac{a_A^{2/3}a_B^{8/3}}{a_C^{10/3}}\left(1+\frac{{\mu_B} a_B^2}{4{\mu_A} a_A^2}\right)^{-2/3},
\label{eq:ecapii}
\end{equation}
 which again requires that the orbit of binary A is nearly circular. 

\label{lastpage}

\end{document}